\documentclass[12pt]{article}

\ifx\pdfoutput\undefined
\usepackage[dvips,bookmarks]{hyperref}
\else
\usepackage{hyperref}
\fi
\hypersetup{colorlinks=false,bookmarksopen,bookmarksnumbered,citecolor=blue,
   pdfstartview=FitH}

\usepackage{latexsym}

\usepackage{amssymb,amsfonts,amsmath}
\usepackage{graphicx}
\usepackage{indentfirst}

\parskip=\medskipamount

\def\a{\alpha}

\def\b{\beta}

\def\X{\Xi}

\newcommand{\sect}[1]{\section{#1}\setcounter{equation}{0}}

\newcommand{\be}{\begin{equation}}
\newcommand{\ee}{\end{equation}}
\newcommand{\bea}{\begin{eqnarray}}
\newcommand{\eea}{\end{eqnarray}}

\newcommand{\ba}{\begin{array}}
\newcommand{\ea}{\end{array}}

\def\double #1{#1{\hbox{\kern-2pt $#1$}}}

\newcommand{\bsubeq}{\begin{subequations}}
\newcommand{\esubeq}{\end{subequations}}

\newcommand{\de}{{\nabla}}

\newcommand{\eq}{\begin{equation}}
\newcommand{\eqa}{\begin{eqnarray}}
\newcommand{\en}{\end{equation}}
\newcommand{\ena}{\end{eqnarray}}
\newcommand{\enn}{\nonumber \end{equation}}

\def\Dcal{{\cal D}}

\def\psibar{\bar \psi}

\def\rhobar{\bar \rho}

\def\thetabar{\bar \theta}

\def\epsi{\varepsilon}

\def\al{\alpha}
\def\be{\beta}
\def\ga{\gamma}

\let \si\sigma
\let \part\partial

\def\unmezzo{{1 \over 2}}
\def\epsi{\varepsilon}

\def\de{\delta}

\def\part{\partial}

\def\X0{X^0}

\def\al{\alpha}
\def\ga{\gamma}

\def\unmezzo{{1 \over 2}}
\def\epsi{\varepsilon}

\def\de{\delta}

\def\Dcal{{\cal D}}

\def\square{{\,\lower0.9pt\vbox{\hrule \hbox{\vrule height 0.2 cm
\hskip 0.2 cm \vrule height 0.2 cm}\hrule}\,}}

\def\Gtilde{\tilde G}

\def\psibar{\bar \psi}

\def\rhobar{\bar \rho}

\def\thetabar{\bar \theta}

\begin{document}
\begin{titlepage}
\begin{flushright}
\par\end{flushright}
\vskip 1.5cm
\begin{center}
\textbf{\huge \bf The Integral Form of Supergravity}

\vskip 1.5cm

{\large
L. Castellani$^{~a,b,}$\footnote{leonardo.castellani@uniupo.it},
R. Catenacci$^{~a,c,}$\footnote{roberto.catenacci@uniupo.it},
and
P.A. Grassi$^{~a,b,}$\footnote{pietro.grassi@uniupo.it}
\medskip
}
\vskip 0.5cm
{
\small\it
\centerline{$^{(a)}$ Dipartimento di Scienze e Innovazione Tecnologica, Universit\`a del Piemonte Orientale} }
\centerline{\it Viale T. Michel, 11, 15121 Alessandria, Italy}
\medskip
\centerline{$^{(b)}$ {\it
INFN, Sezione di Torino, via P. Giuria 1, 10125 Torino} }
\centerline{$^{(c)}$ {\it
Gruppo Nazionale di Fisica Matematica, INdAM, P.le Aldo Moro 5, 00185 Roma} }
\vskip  .5cm
\medskip

\par\end{center}

\vfill{}

\begin{abstract}
{By using integral forms we
derive the superspace action of $D=3, N=1$ supergravity as an integral
on a supermanifold. The construction is based on
target space picture changing operators, here playing the r\^ole of Poincar\'e duals
to the lower-dimensional spacetime surfaces embedded into the supermanifold. We show how
the group geometrical action based on the group manifold approach interpolates between
the superspace and the component supergravity actions, thus providing another proof of their equivalence.}

\end{abstract}
\vfill
\centerline{\it Dedicated to the memory of Mario Tonin.}
\vskip  2cm

\end{titlepage}

\tableofcontents
\newpage
\setcounter{footnote}{0}


\section{Introduction}

Three dimensional supergravity is one of the simplest models of a consistent extension of
general relativity that includes fermions and local supersymmetry. For this reason it has been revisited
as a workable example in many textbooks and research papers
(see for example \cite{GGRS} and \cite{vN,ZP,HT,Uematsu,BG,RuizRuiz:1996mm}. For recent developments see for ex. 
\cite{Becker:2003wb}).  It also provides a
manageable model of superfield supergravity, with a superfield action integrated on superspace.\footnote{We distinguish between {\it superspace} and {\it supermanifold}. The former denotes a flat bosonic spacetime with additional fermionic coordinates, while the latter the full-fledged supermanifold according to \cite{Catenacci:2007lea}.
}  That action (see  \cite{RuizRuiz:1996mm}), supplemented by {\it ad hoc} constraints consistent with the
Bianchi identities, provides an off-shell formulation of $D=3$ supergravity, local supersymmetry being
realized as a diffeomorphism in the fermionic directions.

On the other hand, the construction of a 3d N=1 supergravity in the {\it rheonomic}
(a.k.a. {\it group manifold}) approach\footnote{for reviews on the group manifold approach see for ex.
 \cite{Castellani,Castellani:1981um,Castellani:1992sv} .} provides a superfield action which
yields both the correct spacetime equations of motion, {\it and} the constraints on the curvatures. The action is written as a
Lagrangian 3-form integrated over a bosonic submanifold of the complete supermanifold.
As discussed in \cite{Castellani:2015paa}, the same action
can be written as the integral over the whole supermanifold of an integral form,
using the Poincar\'e dual that encodes the embedding of the 3-dimensional bosonic submanifold.

At the moment, however,
there is no explicit dictionary between the superfield approach and the group manifold approach.

In this paper we find a bridge between the two formalisms by a novel technique based
on the integration of {\it integral} forms. As is well known, differential forms on superspace cannot be
integrated on a supermanifold ${\cal SM}^{(n|m)}$ (where $n$ and $m$ refer to the bosonic and fermionic dimensions,
respectively) since there is no top form in the usual complex of differential forms. Indeed
the fermionic 1-forms behave like commuting variables with respect to the wedge product and
therefore there is no upper bound to the number of fermionic 1-forms. Nonetheless, one can extend the
space of forms by including distribution-like forms (see for example
\cite{Castellani:2015paa,VORONOV1,Catenacci:2007lea,Catenacci:2010cs,Witten:2012bg}). These can
be incorporated into a consistent
differential calculus where top forms do exist, and can be integrated on
the supermanifold.

The bridge between the superspace action of \cite{GGRS,RuizRuiz:1996mm} and the group-manifold formalism
is provided by the group-manifold three-form Lagrangian ${\cal L}$, which is closed (in general
$d {\cal L}=0$ when auxiliary fields are present \cite{Castellani}).
Multiplied by a suitable closed Poincar\'e dual form
(known in the string theory literature as {\it Picture Changing Operator} or PCO) it becomes an integral top form, and therefore can be integrated
on the supermanifold. Choosing  Poincar\'e duals in the same cohomology class does not change
the action if the Lagrangian is closed.

In particular there is a canonical Poincar\'e dual that produces the standard spacetime action with auxiliary fields of
\cite{GGRS,RuizRuiz:1996mm}. Another Poincar\'e dual,
differing from the first by a total derivative, leads to an expression
for the action that coincides with the superfield action  of \cite{GGRS,RuizRuiz:1996mm}. Since the two Poincar\'e duals
are in the same cohomology class, the two actions are equal.

Furthermore, the expression of the action written as the integral of a Lagrangian three-form
times a PCO clarifies an additional issue. As recalled above, the superfield formulation of supergravity is redundant in
the sense that one needs some constraints to limit the number of independent component fields. It would be
advantageous to have the constraints built in directly into the action. This is achieved in the present formulation:
the closure of the PCO implies exactly those constraints.

The paper is organised as follows. In Section 2 we discuss the equivalence between superspace and group-manifold formulations in general terms. In Section 3 we provide the basic ingredients for the superfield and the group manifold formulations of $D=3, N=1$  supergravity: the constraints, the Bianchi identities and their solutions. In Section 4 we prove the equivalence between the group manifold (rheonomic) formulation, the component spacetime action and  the superspace action. In Section 5 we list some perspectives for future work and in the Appendices we give some
further details on the PCO.

\section{Superspace versus supergroup manifold}

We want to formulate $D=3$ $N=1$ supergravity in two frameworks, namely in the group-manifold approach and the superspace approach. Let us first  clarify what we mean by supersymmetric {\it action} in the two frameworks.

\subsection{Superspace}

First, we parametrize
the superspace ${\mathbb R}^{(3|2)}$ with a set of coordinates $(x^a, \theta^\a)$ with $a=1,2,3$ and $\a=1,2$. The same set
of coordinates will be also used to parametrize a local patch of a supermanifold ${\cal SM}^{(3|2)}$.

In the case of superspace (see for example the textbook \cite{GGRS}), the Lagrangian is a superfield
${\cal F}(x,\theta)$, a local functional of the superfields $\phi(x,\theta)$ of the theory. A superfield
can be expanded in its components $\phi_0, \phi_{1,\a}, \phi_2$
\begin{eqnarray}
\label{ssAA}
\phi(x,\theta) = \phi_0(x) + \phi_{1,\a}(x) \theta^\a + \phi_2(x) \frac{\theta^2}{2}\,,
\end{eqnarray}
with $\theta^2 \equiv \epsilon_{\al\be} \theta^\al \theta^\be$ and the components are identified with the physical degrees of freedom. A generic superfield might also
contain some auxiliary fields to complete the spectrum so that there is a match between
off-shell bosonic and fermionic degrees of freedom.

The {\it superspace action} is the functional
\begin{eqnarray}
\label{ssA}
S_{susy}[\phi] = \int[d^3x d^2\theta] {\cal F}(x,\theta)
\end{eqnarray}
where the symbol $[d^3x d^2\theta]$ refers to the integration variables.
The integration over the $\theta$'s is given by the Berezin integral. Varying the action under an infinitesimal deformation
of the superfields $\phi$, we obtain the superfield equations of motion. In the case of supergravity,
the superfields $\phi$ entering (\ref{ssA}) are subject to constraints, and
their variations have to be compatible with these constraints. Given (\ref{ssA}), one can compute
the Berezin integral by expanding the action in powers of $\theta$ and then selecting the highest term
\begin{eqnarray}
\label{ssB}
S_{susy}[\phi]  = \int [d^3x] \left. D^2  {\cal F}(x,\theta)\right|_{\theta=0}
\end{eqnarray}
which is the {\it component action} written in terms of the physical fields. The superderivative is defined
as $D_\a = \partial_\a + (\bar \theta \gamma^a)_\a \partial_a$ where $(\partial_a, \partial_\a)$ are the ordinary derivatives
with respect to $(x^a, \theta^\a)$.  In addition, $D^2 = \epsilon^{\a\b} D_\a D_\b$.

The supersymmetry of the action is
easily checked: since $ {\cal F}(x,\theta)$ is a superfield, its supersymmetry variation is simply
\begin{eqnarray}
\label{ssC}
\delta {\cal F}(x,\theta) = \epsilon^\a Q_\a {\cal F}(x,\theta)\,,
\end{eqnarray}
$Q_\a$ being the supersymmetry generator satisfying the algebra $\{Q_\a, Q_\b\} =
2 i \gamma^a_{\a\b} \partial_a$ where $\gamma^a_{\a\b}$ are the Dirac matrices for $D=3$. The
supersymmetry generator is defined as $Q_\a = \partial_\a - (\bar \theta \gamma^a)_\a \partial_a$

The property (\ref{ssC}) follows from the fact that ${\cal F}(x,\theta)$ is built out of superfields $\phi$,
their derivative $\partial_a$ and superderivative $D_\a$ and products thereof.

In the case of rigid supersymmetry, the action is invariant because the variation
of the Lagrangian is a total derivative.

In the case of local supersymmetry, one needs to impose the vanishing
of $Q_\a {\cal F}(x,\theta)=0$.

 There are several advantages in having a  superspace action as in (\ref{ssA}). It is the most economical and compact
 way to describe the complete action for all physical degrees of freedom of supergravity, it encodes all symmetries,
 it provides a powerful quantization technique, known as {\it supergraph} method,
 which minimises the amount of Feymann diagrams needed for a single scattering amplitude.
 The supersymmetry cancellations  and the non-renormalization theorems are mostly manifest.

 The main drawback of (\ref{ssA}) is the lack of a fully geometrical interpretation,
 since it cannot be understood as an integral of a differential form on a manifold. The expression for the superfield action is
 usually dictated by scaling properties and Lorentz covariance, but it is not very intuitive and for constrained
 superfields it does not always exist. In that respect the group-manifold approach seems to overcome these problems.

 \subsection{Supergroup manifold}

The logic of this approach is algebraic: one starts from a superalgebra, and to each generator $T_A$ corresponds
a one-form (vielbein) field $\sigma^A$ on the supergroup manifold $G$. The vielbein satisfies the Cartan-Maurer equations:
 \eq
  d \sigma ^A + \unmezzo C^{A}_{~BC} \sigma^B \wedge \sigma^C =0
   \en

The fields of the theory are identified with the various components of the vielbein $\sigma^A$, labelled by the
adjoint index {\it \small A}. For the fields to become dynamical, they must be allowed to develop a nonzero curvature, that is to say
the right-hand side of the Cartan-Maurer equations must be nonvanishing in general. This is achieved by considering
deformations of the supergroup manifold, i.e. a ``soft" supergroup manifold $\Gtilde$.

A systematic procedure \cite{Castellani,Castellani:1981um,Castellani:1992sv} leads to the construction of $d$-form
lagrangians, whose restriction to a $d$-dimensional
bosonic manifold reproduces the $d$-dimensional spacetime supergravity lagrangians. The local symmetries of the theory
are the superdiffeomorphisms on $\Gtilde$, and include the supersymmetries as diffeomorphisms in the fermionic
directions of $\Gtilde$. In this respect supersymmetry transformations have a geometric interpretation similar
to the one in the superfield approach.

The supervielbein field $\sigma^A$ is a 1-superform living in $\Gtilde$. The coordinates of $\Gtilde$ are the spacetime coordinates $x^\mu$,
corresponding typically to the translation subgroup of $G$, Grassmann coordinates $\theta^\alpha$, corresponding
to the fermionic generators of $G$, and other coordinates corresponding to gauge directions. Diffeomorphisms in these
last coordinates produce gauge transformations, and the dependence of the fields on these coordinates can be removed via
a finite gauge transformation. At the end of the game all fields depend on $x$ and $\theta$.

Still one has a great redundancy, since $\sigma^A$ is expanded as a superspace 1-form as
 \eq
  \sigma^A (x, \theta) =  \sigma^A (x,\theta)_a dx^a + \sigma^A (x,\theta)_\a d\theta^\a
   \en

Typically the fields one wants to retain as dynamical fields in this formulation are given by $\sigma^A (x, \theta=0)$.
In other words one has to eliminate the extra degrees of freedom due to the $\theta$ dependence and to the $d \theta$ components.

The variational principle involves variations of the fields, and variations of the embedding in $\Gtilde$ of the bosonic submanifold.
The resulting equations yield the usual spacetime field equations, together with the constraints needed to remove the
redundant degrees of freedom (``rheonomic constraints").

In terms of these ingredients, the $D=3$, $N=1$ {\it rheonomic action} is defined as the integral over a bosonic submanifold ${\cal M}^{(3)}$
of the supermanifold  ${\cal SM}^{(3|2)}$ as follows
\begin{eqnarray}
\label{ssD}
S_{rheo}[\sigma, {\cal M}^{(3)}] = \int_{ {\cal M}^{(3)} \subset {\cal SM}^{(3|2)}} {\cal L}^{(3)}(x,\theta,dx, d\theta)
\end{eqnarray}
and depends on the superforms $\si$ and on the embedding of ${\cal M}^{(3)}$ into the supermanifold
${\cal SM}^{(3|2)}$.
Changing the embedding corresponds to a diffeomorphism and it can be compensated by a change of the
Lagrangian $  {\cal L}^{(3)}(x,\theta,dx, d\theta)$, generated by a Lie derivative. Therefore the variational
equations can be obtained by varying the fields for an arbitrary embedding, and considering the resulting
equations as 2-form equations on the whole superspace.  Projections of these equations in the fermionic directions
($d\theta$ directions) yield
the rheonomic constraints, necessary to remove unwanted degrees of freedom. The  correct component action
is retrieved by setting $\theta =0$ and $d\theta =0$  (see the textbook \cite{Castellani}) .

The supersymmetry of the action is
expressed as a diffeomorphism in the fermionic directions of the supermanifold and therefore the variation of ${\cal L}^{(3)}$ is
given by
\begin{eqnarray}
\label{ssE}
\delta {\cal L}^{(3)} = \ell_\epsilon {\cal L}^{(3)} = d (\iota_\epsilon {\cal L}^{(3)}) + \iota_\epsilon d {\cal L}^{(3)}\,.
\end{eqnarray}
If the Lagrangian satisfies
 \eq
   \iota_\epsilon d {\cal L}^{(3)}=0 \label{idL}
   \en
 the variation of ${\cal L}^{(3)}$ is a total derivative and the action is invariant. Condition (\ref{idL}) is
 in fact equivalent to the rheonomic constraints mentioned above.

 The form of ${\cal L}^{(3)}$ has a direct correspondence with the component action, to which it reduces
 after setting $\theta=0$ and $d \theta=0$. It is less compact than the superfield formulation, but
 more transparently related to the component action.

 We have argued that the local symmetries of the group manifold
 action are the diffeomorphisms on the supergroup manifold. This certainly holds true if one considers a group manifold
 action resulting from the integration of a {\it top} form on $\Gtilde$. Since the past literature on group manifold
 actions for supergravity makes little reference to superintegration theory, this point has needed some clarification,
 reported in \cite{Castellani:2015paa,Castellani:2014goa}, and involves Poincar\'e duals and integral top forms.

 \subsection{Equivalence}

The component action obtained in the two formulations must be the same or, at least,
 related by field redefinitions. Therefore there must exist a {\it mother} action
 which interpolates between the two formulations. This action is the rheonomic action.
 The way to
 integrate a $3$-form on a submanifold of a bigger manifold is by constructing a
 Poincar\'e  dual of that submanifold, and denoting it by ${\mathbb Y}^{(0|2)}$ the supersymmetric action is
 given by
 \begin{eqnarray}
\label{ssF}
S_{susy}[\phi] = \int_{{\cal SM}^{(3|2)}} {\cal L}^{(3|0)} \wedge {\mathbb Y}^{(0|2)}
\end{eqnarray}
where ${\cal L}^{(3|0)}$ is the rheonomic Lagrangian used in (\ref{ssD}) and the integration is on the complete supermanifold. The
Poincar\'e dual (also known as PCO) localizes the full supermanifold to the submanifold.
Integration on supermanifolds is discussed in several papers (see for example \cite{Castellani:2015paa} for the definition of the Poincar\'e dual on supermanifolds). Only the integral forms can be integrated. The complex of
differential forms on a supermanifold contains the pseudo-forms which are polynomials in $dx^a, d\theta^\a,
\delta^{(p)}(d\theta^\alpha)$ (where $\delta^{(r)}$ are $r$-derivatives of the delta function) .
They are characterized by two numbers $(p|q)$: the {\it form degree p} and the {\it picture number q} where the latter
counts the number of delta functions. In general $(p|q)$-forms are integral forms on ${\cal SM}^{(p|q)}$,
and can be integrated on this supermanifold.
The integral forms of ${\cal SM}^{(3|2)}$ are those with $(3|2)$ and they can be integrated on
${\cal SM}^{(3|2)}$. Thus the Lagrangian ${\cal L}^{(3|0)} \wedge {\mathbb Y}^{(0|2)}$ is an integral form, built as the product of
the rheonomic action ${\cal L}^{(3|0)}$, which is a $(3|0)$-superform (constructed as discussed above), and
the Poincar\'e dual/PCO ${\mathbb Y}^{(0|2)}$, which is a $(0|2)$-form.

The Poincar\'e dual/PCO ${\mathbb Y}^{(0|2)}$ is closed and not exact (it belongs to the cohomology class
$H^{(0|2)}(d, {\cal SM}^{(3|2)})$), and its variation under the change of the embedding of ${\cal M}^{(3)}$ into
${\cal SM}^{(3|2)}$ is $d$-exact:
\begin{eqnarray}
\label{ssG}
\delta {\mathbb Y}^{(0|2)} = d \Omega^{(-1|2)}
\end{eqnarray}
where  $\Omega^{(-1|2)}$ is an integral form with negative form degree
(derivatives of the delta functions act as negative degree forms: for example $d\theta \delta'(d\theta) = - \delta(d\theta)$).
Then, any variation of the embedding is ineffective
if  ${\cal L}^{(3|0)}$ is closed (the action does not depend on the embedding). Also, if two ${\mathbb Y}$'s are related by $d$-exact terms,
namely if they belong to the same cohomology class,  the corresponding actions are equivalent.

We propose the two different choices
\begin{eqnarray}
\label{ssH}
{\mathbb Y}^{(0|2)}_{st} = \theta^2 \delta(d\theta)\,, ~~~~~~~~~~~~~~~~
{\mathbb Y}^{(0|2)}_{susy} = V^a \wedge V^b \gamma_{ab}^{\a\b} \iota_\a \iota_\b \delta^2(\psi)\,,
\end{eqnarray}
where $(V^a, \psi^\a)$ are the components of the supervielbein $E^A$.
$\iota_\a$ is the derivative of the delta function with respect to its argument and
$ \delta^2(\psi) = \epsilon_{\a\b} \delta(\psi^\a) \wedge \delta(\psi^\b)$. Inserting the first
PCO ${\mathbb Y}^{(0|2)}_{st}$ we project the Lagrangian to ${\cal L}^{(3|0)}(x,0,dx,0)$ yielding
the component action. The second choice leads to the superspace action in (\ref{ssA}). The main goal
of the present work is to prove this equivalence.

In a related work \cite{grassi-mac}, the equivalence of the different formulations
of $N=1$ super Chern-Simons theory has been studied. The flat version of ${\mathbb Y}^{(0|2)}_{susy}$ is discussed and
its properties are described in that paper.

\sect{$D=3$, $N=1$ supergravity in the two frameworks}

The theory contains a vielbein 1-form $V^a$ with 3 off-shell degrees of freedom ($d(d-1)/2$ in $d$ dimensions),
and a gravitino $\psi^\a$ with 4 off-shell degrees of freedom ($(d-1)2^{[d/2]}$ in $d$ dimensions for Majorana or Weyl).
The mismatch can be cured by an extra bosonic d.o.f., here provided by a bosonic 2-form auxiliary field $B$.
As recalled, the group-geometric procedure to build supergravity actions starts from a superalgebra. In the case at hand the superalgebra is the superPoincar\'e algebra,
generated by $P_a, L_{ab}$ and $Q_\a$ (the translation generators, the Lorentz generators and the supersymmetry charges).
The structure constants of the superalgebra are encoded in the Cartan-Maurer equations
 \eq
  d \sigma^A + \frac12 C^A_{~BC} \sigma^B \wedge \sigma^C = 0
  \en
  \noindent where the left-invariant one-forms $\sigma^A$ are a cotangent (vielbein) basis,
  dual to the tangent vectors on the supergroup manifold $G$. In the present case the cotangent basis is given
  by the vielbein $V^a$, the spin connection $\omega^{ab}$ and the gravitino $\psi^\al$. The algebra is further extended with a 2-form $B$ in order
  to match the degrees of freedom (and thus becomes a {\it Free Differential Algebra} (FDA),
  see for ex. \cite{Castellani}).

    The generalized Cartan-Maurer equations of the FDA yield the definitions of the Lorentz curvature, the torsion, the gravitino field strength and the 2-form field strength:
   \eqa\label{parB}
  & & R^{ab}=d \omega^{ab} - \omega^a_{~c} ~ \omega^{cb} \\
   & & R^a=dV^a - \omega^a_{~b} ~ V^b - {i \over 2} \psibar \gamma^a \psi \equiv \Dcal V^a - {i \over 2} \psibar \gamma^a \psi\ \label{torsionRa}\\
   & & \rho = d\psi - {1 \over 4} \omega^{ab} \gamma_{ab} ~ \psi \equiv  \Dcal \psi \\
   & & H=dB-{i \over 2} \psibar \gamma^a \psi ~V^a    \ena
where $\Dcal$ is the Lorentz covariant derivative, and exterior products between forms are understood.
The Cartan-Maurer equations are invariant under rescalings
     \eq
     \omega^{ab} \rightarrow \lambda^0 \omega^{ab}, ~V^a \rightarrow \lambda  V^a,~\psi \rightarrow \lambda^{1\over 2} \psi,~B \rightarrow \lambda^2 B \label{rescalings}
           \en
     Taking exterior derivatives of both sides yields the Bianchi identities:
     \eqa\label{parC}
    & &  \Dcal R^{ab} =0 \\
    & &  \Dcal R^a + R^a_{~b} ~ V^b - i~ \psibar \gamma^a \rho =0\\
    & & \Dcal \rho + {1 \over 4} R^{ab} \gamma_{ab} ~\psi =0\\
    & & dH- i~ \psibar \gamma^a \rho V^a + {i \over 2} \psibar \gamma^a \psi~ R^a = 0
     \ena
     invariant under the rescalings (\ref{rescalings}).

As explained above, the redundancy introduced by promoting each physical field to a superfield has to be tamed
by imposing some algebraic constraints on the curvature parametrizations. They are known as {\it conventional constraints} in the superspace language and as {\it rheonomic parametrizations} in the group-manifold approach. We use
the following parametrizations
       \eqa\label{parA}
         & & R^{ab} = R^{ab}_{~~cd} ~V^c V^d + \thetabar^{ab}_{~~c}~\psi ~V^c + c_1~ f ~\psibar \gamma^{ab} \psi \\
        & & R^a = 0  \label{parRa}\\
        & & \rho = \rho_{ab} V^a V^b + c_2~f~\gamma_a \psi ~V^a \\
        & & H = f~V^a V^b V^c \epsilon_{abc} \\
        & & df = \partial_a f~ V^a + \psibar \Xi \label{pardf}
        \ena

        with
         \eqa
         \thetabar^{ab}_{~~c, \a} = c_3 ~(\bar\rho_c^{~[a} \gamma^{b]})_\a + c_4 (\rhobar^{ab} \gamma_c)_\a~~~~~~~~~~
         \Xi_\a =c_5 ~ \epsilon^{abc} (\gamma_a \rho_{bc})_\a
         \ena
         The coefficients $c_1,c_2, c_3, c_4,c_5$ are fixed by the Bianchi identities
        to the values:
         \eq
         c_1=  {3i \over 2} ,~c_2= {3 \over 2} ,~c_3 = 2i,~c_4=-i,~c_5 = - {i \over 3!}
         \en

         The $VVV$ component $f$ of $H$ scales as $f \rightarrow \lambda^{-1} f$,
         and is identified with the auxiliary scalar superfield of the superspace approach of ref \cite{RuizRuiz:1996mm}.
         Note that, thanks to the presence of the auxiliary field, the Bianchi identities do not imply
         equations of motion for the spacetime components of the curvatures. To compare with the superspace
         approach and the superspace action, we have to clarify the role of the superfield $f$.

The superspace formulation of supergravity in $D=3$ follows a different path, and considers the supervielbein $E^A$ and
the spin connection $\omega^A_{~B}$ as fundamental fields, with {\small A=a, $\al$}. The index of the supervielbein now runs
only on the superspace directions, and $E^A$ contains the fields of the rheonomic approach as $E^a = V^a$, $E^\al = \psi^\al$.

 Again there is a huge redundancy in that formulation,
and one has to impose some constraints. First, one imposes the {\it soldering} constraint on the
spin connection
\begin{eqnarray}
\label{ciccA}
\omega^A_{~B} =
\left(
\begin{array}{ccc}
 \omega^a_{~b} & 0    \\
 0 & \frac14 (\gamma^{ab})^{\a}_{~\b} \omega_{ab}
\end{array}
\right)\,,
\end{eqnarray}
 where the off-diagonal pieces are set to zero and the spinorial part of the connection is
 related to the Lorentz spin connection. As a consequence the supercurvature
 	\eq
	R^{AB} = d \omega^{AB} - \omega^A_{~C} \wedge \omega^{CB}
	\en
has nonvanishing components $R^{ab}$, $R^{\al\be} =  \frac14 (\gamma_{ab})^{\a}_{~\b} R^{ab}$ with
 \begin{eqnarray}
\label{}
R^{ab} = R^{ab}_{rs} E^r \wedge E^s + R^{ab}_{r\sigma} E^r \wedge E^\sigma + R^{ab}_{\rho\sigma} E^\rho \wedge E^\sigma\,,
\end{eqnarray}
The superfields
$R^{ab}_{rs},  R^{ab}_{r\sigma}$ and $R^{ab}_{\rho\sigma}$ correspond to the analogous terms in (\ref{parA}).
Similarly,
one considers the supertorsion
\eq
 {T^A} = d E^A - \omega^A_{~B} \wedge E^B
 \en
  which has the following
expansion on the supervielbein basis
\begin{eqnarray}
\label{ciccB}
T^a &=&  T^{a}_{~rs} E^r \wedge E^s + T^{a}_{~r\sigma} E^r \wedge E^\sigma + T^{a}_{~\rho\sigma} E^\rho \wedge E^\sigma\,, \nonumber \\
T^\a &=&  T^{\a}_{~rs} E^r \wedge E^s + T^{\a}_{~r\sigma} E^r \wedge E^\sigma + T^{\a}_{~\rho\sigma} E^\rho \wedge E^\sigma \,.
\end{eqnarray}
 To reduce the
independent components, one imposes the {\it conventional constraints}
\begin{eqnarray}
\label{ciccC}
T^a_{~\rho\sigma}  = \frac12 i \gamma^a_{\rho\sigma}\,, ~~~~~
T^a_{~ r \sigma} =0\,, ~~~~~
T^\a_{~ \rho\sigma} = 0\,, ~~~~
T^\a_{~ r\sigma} = 2 i (\gamma_r)^\a_{~\sigma} R\,, ~~~~
\end{eqnarray}
The Bianchi identities then imply $R^{ab}_{~\rho\sigma} =0$ and $T^a_{~rs} =
\epsilon^a_{~rs} R$, where $R$ is a superfield containing the scalar auxiliary field as first component, the
gravitino curvature as mixed component, and  the Ricci scalar as  $\theta^2$ component.
The solution for the other components can be found in \cite{RuizRuiz:1996mm,Kuzenko:2011xg}.
The supertorsion $T^a$ differs from $R^a$ defined in (\ref{torsionRa}) by a term bilinear in fermions,
and this reflects into the first constraint given above.

Using these constraints, one finds that the only independent off-shell degrees of freedom (vielbein, gravitino and scalar auxiliary field)
 are contained in the components
$E^\a_\mu$ and $E^\a_m$ of the superform expansion $E^\a = E^\a_\mu d\theta^\mu + E^\a_m dx^m$. Using the
gauge symmetries, one can identify the physical and auxiliary fields.

Comparing the analysis in the superspace and the
analysis in the rheonomic approach, we find that the auxiliary superfield $f$ has to be identified with $R$. Indeed
we observe that, by a change of the spin connection, one can set to zero the last term in the parametrization of the curvature $R^{ab}$ in (\ref{parA}),  namely $R^{ab}_{~\rho\sigma} =0$.  This change in the spin connection produces a change  of $R^a$
in (\ref{parRa}) such that $R^a_{~rs} = \epsilon^a_{~rs} f$. Comparing with the constraint   $T^a_{~rs} =
\epsilon^a_{~rs} R$ of the superfield approach one finds $f = R$.

\sect{The actions and their equivalence}

To uncover the relation between the superspace action (\ref{ssA}), the rheonomic action (\ref{ssD}) and the component action, we have to discuss them in the corresponding frameworks.

         With the usual group-geometrical methods, the action is determined as in (\ref{ssD}) and the Lagrangian
     ${\cal L}^{(3)}$ reads
          \eq
 {\cal L}^{(3)} =  R^{ab} V^c \epsilon_{abc} + 2i \psibar \rho + \alpha (f H - {1 \over 2} f^2 V^a V^b V^c \epsilon_{abc})
           \label{spacetimeaction}
            \en
            This action is obtained by taking for the Lagrangian ${\cal L}^{(3)}$ the most general Lorentz scalar 3-form, given in terms of the curvatures
            and 1-form fields (cotangent basis of ${\tilde G}$), invariant under the rescalings discussed above, and then
            requiring that the variational equations admit the vanishing curvatures solution
             \eq
             R^{ab} = R^a = \rho=H = f = 0\,,
             \en
             and also imply the constraints, arising from the $\delta \omega^{ab}$ and $\delta f$ variations:
              \eq
               R^a =0, ~~~H = f \epsilon_{abc} V^a V^b V^c \,.
               \en
               The remaining parameter $\alpha$ is fixed by requiring the closure of ${\cal L}^{(3)}$ , i.e. $d{\cal L}^{(3)} =0$.
               This yields $\alpha = 6$, and ensures the off-shell closure of the supersymmetry transformations
               given below. The action is invariant under off-shell supersymmetry transformations which are
               easily computed by taking the Lie derivative of the fields along
               the fermionic directions (tangent vectors dual to $\psi^\a$):
                \eqa
              & &  \delta_\epsi V^a = -i \psibar \gamma^a \epsi \\
              & &  \delta_\epsi  \psi = \Dcal \epsi \\
              & & \delta \omega^{ab}= \thetabar^{ab}_{~~c} ~ \epsi V^c - 3i f~ \psibar \gamma^{ab} \epsi \\
              & & \delta_\epsi  B = - i \psibar \gamma^a \epsi V^a \\
              & & \delta_\epsi f = 0
                \ena
               and close on all the fields without need of imposing the field equations.

                       Varying $\omega^{ab}$, $V^a$, $\psi$, $B$ and $f$ leads to the equations of motion:
                \eqa \label{reoEQ}
                 & & R^a=0 \\
                 & & R^{ab} = 9 f^2 V^a V^b + {3i \over 2} f ~ \psibar \gamma^{ab} \psi \\
                 & & \rho = {3 \over 2} \gamma_a \psi ~ V^a \\
                 & & df=0 \\
                 & & H= f~V^a V^b V^c \epsilon_{abc}
                 \ena

            Notice that the equations of motion are obtained from the rheonomic action principle (as explained in
            the textbook \cite{Castellani}), by varying the action keeping the submanifold fixed. They are 2-form equations and
             can be expanded on the basis $V^a, \psi^\a$.

Let us move to the superspace action. As we have seen in the previous Section, after imposing the constraints we are left with
a superfield $R$ which contains the auxiliary field, the Ricci scalar and the Rarita-Schwinger term. To build the action we therefore
consider the expression
\begin{eqnarray}
\label{ciccE}
{\cal F}(x,\theta) = R \, {\rm Sdet}(E)
\end{eqnarray}
where ${\rm Sdet}(E)$ is the superdeterminant of the supervielbein $E^A$. The expression in ${\cal F}(x,\theta)$ is
a superfield and transforms as discussed in Sec. 2. By expanding at the second order in $\theta$'s, one can retrieve the
component action. However, the computation is rather cumbersome already in the present simplified context. A better way to
derive the component action from (\ref{ciccE}) is the use of the {\it ectoplasmic} integration theory
\cite{Gates:1997kr,Gates:1998hy,Gates:2009uv,Kuzenko:2013uya}. We refer to
\cite{GGRS,RuizRuiz:1996mm} for a complete discussion and for the equations of motion in superspace.

Finally, we are ready to discuss the relation between the two actions. As explained in the introduction,
the {\it mother} theory interpolating between the rheonomic action, the superspace action and the
component action is described by the superintegral:
\begin{eqnarray}
\label{intA}
S_{SG} = \int_{{\cal SM}^{(3|2)}} {\cal L}^{(3|0)} \wedge {\mathbb Y}^{(0|2)}
\end{eqnarray}
where the Lagrangian ${\cal L}^{(3|0)}$ is the rheonomic action given in (\ref{spacetimeaction}). It is a $(3|0)$-form and
it is closed because of the parametrizations (\ref{parA})-(\ref{pardf}).\footnote{The dependence of the fields on the gauge (Lorentz) coordinates factorizes, and reduces to a multiplicative factor in front of the integral over the superspace.} The choice of the
Poincar\'e dual/PCO ${\mathbb Y}^{(0|2)}$  allows us to interpolate between the component action and the superspace action.

 To retrieve the usual spacetime action one chooses for the Poincar\'e dual/PCO the following $(0|2)$-form:
                 \eq
                 {\mathbb Y}^{(0|2)}_{st} = \epsilon_{\al\be} \theta^\al \theta^\be~\epsilon_{\ga\de} \delta (d\theta^\ga) \delta (d\theta^\de) \label{dtheta2}
                 \en
It is closed and not exact, and it is an element of the cohomology $H^{(0|2)}(d, {\cal SM}^{(3|2)})$.
The integration over the $d\theta$'s is performed by integrating on the Dirac delta functions, that imposes $d\theta =0$.
 Berezin integration in (\ref{intA})  yields an ordinary spacetime action, integrated on ${\cal M}^{(3)}$:
                    \eq
                   S_{SG} =    \int_{{\cal M}^{(3)}} {\cal L}^{(3|0)} (\theta=0, d\theta=0)
                     \en
                      where all forms depend only on $x$ because of the two $\theta$'s in  ${\mathbb Y}^{(0|2)}_{st}$.  Notice that
                      the supersymmetry variation of ${\mathbb Y}^{(0|2)}_{st}$ is not zero, but is exact, and therefore
                      the integrand is supersymmetric only up to a total derivative.

%

                   The action (\ref{intA}) depends in general on the choice of the bosonic $M^3$ submanifold.
               This choice is encoded in the Poincar\'e dual/PCO ${\mathbb Y}^{(0|2)}_{st}$.
               Varying the submanifold via a diffeomorphism in the
               $\theta$ directions corresponds to a variation of ${\mathbb Y}^{(0|2)}_{st}$ given by an exact form, since the Lie 	derivative
               ${\cal L}_\epsilon = i_\epsilon d + d i_\epsilon$ applied on ${\mathbb Y}^{(0|2)}_{st}$ yields
               $d ( i_\epsilon {\mathbb Y}^{(0|2)}_{st})$. Then the
               variation of the action due to the variation of the submanifold is:
                \eq
                  \delta S_{SG} =  \int_{{\cal SM}^{3|2}} {\cal L}^{(3|0)} \wedge d ( i_\epsilon {\mathbb Y}^{(0|2)}_{st})
                    \en
                Integrating by parts and noting that $0=i_\epsilon (d {\cal L}^{(3|0)} \wedge {\mathbb Y}^{(0|2)}_{st})$
                since $d {\cal L}^{(3|0)} \wedge {\mathbb Y}^{(0|2)}_{st}=0$ (because it exceeds the
                maximal rank of an integral form), we find that $\delta S_{SG} =0$  if
                 \eq
                   i_\epsilon d {\cal L}^{(3|0)}=0
                    \en

%
Another
Poincar\'e dual can be chosen as follows
\eq
                        {\mathbb Y}^{(0|2)}_{susy} =  V^a V^b  \gamma_{ab}^{\al\be} i_\al i_\be \delta^2 (\psi)
                         \en
                     with
                      \eq
                       i_\al \equiv {\delta \over \delta \psi^\al},~~~\delta^2 (\psi) \equiv \epsilon_{\ga\de} \delta (\psi^\ga) \delta (\psi^\de)
                        \en
                      We prove in the Appendix that ${\mathbb Y}^{(0|2)}_{susy}$ is connected to the Poincar\'e dual/PCO
                      in (\ref{dtheta2}) by a
                      $\theta$-diffeomorphism. Therefore their difference is exact (since a Lie derivative acting on a closed form
                      gives an exact form), and we find the equivalence:
                       \eq
                        S_{SG} = \int_{{\cal SM}^{(3|2)}} {\cal L}^{(3|0)} \wedge
                        {\mathbb Y}^{(0|2)}_{st} = \int_{{\cal SM}^{(3|2)}} {\cal L}^{(3|0)} \wedge {\mathbb Y}^{(0|2)}_{susy}
                        \en
                        since $d {\cal L}^{(3|0)}=0$. The choice of ${\mathbb Y}^{(0|2)}_{susy}$ is also dictated by Hodge duality:
                        indeed it is the Hodge dual of the $(3|0)$-form:
                         \eq
                          \psibar \gamma_a \psi V^a
                           \en
                           which is closed (by the 3d Fierz identity) and not exact. Since Hodge duality maps $(3|0)$-cohomology classes
                            into $(0|2)$-cohomology classes
                           \cite{Castellani:2015ata}
                           we know a priori that ${\mathbb Y}^{(0|2)}_{susy}$ is closed and not exact, and fulfills the requirements for
                           a Poincar\'e dual.

                        Computing now the term with ${\mathbb Y}^{0|2}_{susy}$, we see that only the first two terms of
                        ${\cal L}^{(3|0)}$ contribute, and using
                        the curvature parametrizations for $R^{ab}$ and $\rho$ one finds:
                         \eq
                          S_{SG} = 6i \int_{{\cal M}^{(3|2)}}  f \epsilon_{abc} V^a V^b V^c \delta^2 (\psi) = 6i \int [d^3x d^2\theta]
                          f {\rm Sdet} (E)
                           \en
                           where $E=(V^a,\psi^\al)$ is the supervielbein in superspace and we have used
\begin{eqnarray}
\label{volume}
{\rm Vol}^{(3|2)} = \epsilon_{abc} V^a \wedge  V^b \wedge V^c \wedge \delta^2 (\psi) = {\rm Sdet}(E) d^3x \delta^2(d\theta)
\end{eqnarray}
            Recalling that $f$ is identified with the scalar superfield $R$ we finally conclude that the two actions are indeed equivalent.\footnote{The relation between integral forms and superspace formulation has also been used
          to formulate massive supergravity in the multivielbein formulation in \cite{grassi-francia}}

                           The present formulation permits also the introduction of a cosmological constant term. This is
                           achieved by shifting the superfield $f$ by a constant term $f \rightarrow f + \sqrt{\Lambda}$ or
                           equivalently, in the superspace framework, by shifting the superfield $R$. The
                           result is that the action acquires a new term proportional to the volume form ${\rm Vol}^{(3|2)}$.
                           It is interesting to notice that this new term
                           \begin{eqnarray}
\label{volumeB}
\Omega^{(3|2)} = \Big( \sqrt{\Lambda} \bar E \gamma_a E E^a - 6i \Lambda \epsilon_{abc} E^a E^b E^c \Big)\,,
\end{eqnarray}
 is closed using the rheonomic parametrizations (\ref{parA})-(\ref{pardf}).

                           In conclusion, the group-manifold rheonomic Lagrangian ${\cal L}^{(3|0)}$,
                           integrated on superspace, yields
                           both the
                           usual spacetime $D=3$ and $N=1$ supergravity action, and its superspace version. The essential ingredients of the proof are
                           Poincar\'e duals differing by a total derivative, and the rheonomic constraints with the auxiliary field that ensure
                           $d {\cal L}^{(3|0)}=0$.

\section{Outlook and Perspectives}

With the present work, we have established a precise mathematical relation between two different superspace
formulations of supergravity. We have used the $N=1, D=3$ supergravity for simplicity. Nonetheless, the present formulation
is applicable to any
supergravity model and in particular to
$N=1,2,4, \dots$ $D=4$ supergravity and higher dimensional models. The mathematical framework permits
 to explore different choices of $PCO$ interpolating different superspace formulations.
An important remark: the equivalence between the different formulations holds because the Lagrangian
${\cal L}^{(3|0)}$ is closed, and this is a consequence of the existence of the auxiliary fields for the model at hand,
i.e. the existence of an off-shell formulation of the theory.
This agrees with the common belief about the existence of an action principle in superspace. Note however that
the rheonomic formulation of supergravity models (such as for example $D=11$ and $D=10$ $N=2$ supergravities)
is available even in absence of auxiliary fields and it would certainly be interesting  to test the present analysis on such models.

 As a final comment we observe that the form ${\cal L}^{(3|0)}\wedge {\mathbb Y}^{(0|2)}$ is
 integrable on the supermanifold ${\cal SM}^{(3|2)}$, but is definitely not the only one. One can
wonder whether it would be possible to construct a supergravity action as a non-factorized $(3|2)$ integral form
 \begin{eqnarray}
\label{ouA}
S_{SG} = \int_{{\cal SM}^{(3|2)}} {\cal L}^{(3|2)}
\end{eqnarray}
 where ${\cal L}^{(3|2)} = \sum_{l=0}^2 {\cal L}^{(3|l)}\wedge {\mathbb Y}^{(0|2-l)}$. Indeed, it can be shown that
 in $N=1$ $D=3$ case, there exists such a possibility and it will be discussed separately.

\section*{Acknowledgements}
We would like to thank C. Maccaferri, D. Francia, F. Del Monte, P. Fr\'e and M. Porrati for useful
discussions and remarks.

\section{Appendix: Properties of the susy PCO}

 \subsection*{Closure}
 The closure of ${\mathbb Y}_{susy}^{(0|2)}$ may be inferred by Hodge duality with the
 cohomology class $\psibar \gamma_a \psi V^a$. In this Appendix we prove it directly.
 We use here the superspace notations for the supervielbein $E^a = V^a, E^\al=\psi^\al$. We check that, by using the conventional supergravity constraints, the PCO \begin{eqnarray}
\label{LCSB}
{\mathbb Y}^{(0|2)} = E^a \wedge E^b (\gamma_{ab})^{\a\b} \iota_\alpha \iota_\beta \delta^2(E)
\end{eqnarray}
is closed and not exact where $\delta^2(E) \equiv \epsilon_{\rho\sigma}\delta(E^\rho) \delta(E^\sigma)$.

It is invariant under Lorentz symmetry since all tangent indices are contracted with Lorentz invariant tensors. It is also
closed. To prove it, we observe
\begin{eqnarray}
\label{LCSC}
&&d \Big( E^a \wedge E^b (\gamma_{ab})^{\a\b} \iota_\alpha \iota_\beta
\left( \epsilon_{\rho\sigma}\delta(E^\rho) \delta(E^\sigma) \right)
\Big) \nonumber \\
&&= 2 \Big( T^a \wedge E^b (\gamma_{ab})^{\a\b} \iota_\alpha \iota_\beta
\left( \epsilon_{\rho\sigma}\delta(E^\rho) \delta(E^\sigma) \right) \Big)\nonumber \\
&&+ 2 \Big( E^a \wedge E^b (\gamma_{ab})^{\a\b} \iota_\alpha \iota_\beta
\left( \epsilon_{\rho\sigma}\iota_\gamma \delta(E^\rho)\wedge T^\gamma\wedge \delta(E^\sigma) \right)\Big)
\end{eqnarray}
We expand the torsion $T^A$ in the vielbein basis: $T^A = T^A_{~BC} E^B\wedge E^C$ and we obtain for the first term:
\begin{eqnarray}
\label{LCSD}
&&T^a \wedge E^b (\gamma_{ab})^{\a\b} \iota_\alpha \iota_\beta
\left( \epsilon_{\rho\sigma}\delta(E^\rho) \delta(E^\sigma) \right) \nonumber \\
&&=
(T^a_{~cd} E^c \wedge E^d  + T^a_{\delta \gamma} E^\delta \wedge E^\gamma)
\wedge E^b (\gamma_{ab})^{\a\b} \iota_\alpha \iota_\beta
\left( \epsilon_{\rho\sigma}\delta(E^\rho) \delta(E^\sigma) \right)
\nonumber \\
&&=
({R} \,\epsilon^a_{~c d} E^c \wedge E^d  + 2 i \gamma^a_{\delta \gamma} E^\delta \wedge E^\gamma)
\wedge E^b (\gamma_{ab})^{\a\b} \iota_\alpha \iota_\beta
\left( \epsilon_{\rho\sigma}\delta(E^\rho) \delta(E^\sigma) \right)
\end{eqnarray}
where we have used the parametrization of the torsion. Due to antisymmetrization, we can recast the first term
as follows
\begin{eqnarray}
\label{LCSE}
{R}\, \epsilon^a_{~c d} \epsilon^{c d b} E^3 (\gamma_{ab})^{\a\b} \iota_\alpha \iota_\beta
\left( \epsilon_{\rho\sigma}\delta(E^\rho) \delta(E^\sigma) \right)  = 0
\end{eqnarray}
where $E^3 = \frac{1}{3!}  \epsilon_{a b c} E^a \wedge E^b \wedge E^c$,
$\epsilon^a_{~c d} \epsilon^{c d b} = \eta^{ab}$, and the term vanishes because of the antisymmetry of
$\gamma_{ab}$. The second term in (\ref{LCSD}) can be written as
\begin{eqnarray}
\label{LCSF}
2 i \gamma^a_{\delta \gamma} (\gamma_{ab})^{\a\b}  \iota_\a E^\delta \wedge \iota_\beta E^\gamma
\wedge E^b \epsilon_{\rho\sigma}\delta(E^\rho) \delta(E^\sigma) = 0
\end{eqnarray}
where we have used  $\iota_\a E^\b = \delta_\a^{~\b}$ by definition.

Let study the second piece in (\ref{LCSC})
\begin{eqnarray}
\label{LCSG}
&&E^a \wedge E^b (\gamma_{ab})^{\a\b} \iota_\alpha \iota_\beta \left[
\epsilon_{\rho\sigma}\iota_\gamma \delta(E^\rho)\wedge \Big(T^\gamma_{cd} E^c \wedge E^d +
T^\gamma_{c \delta} E^c \wedge E^\delta\Big) \wedge \delta(E^\sigma) \right] \nonumber \\
&&=
E^a \wedge E^b  \wedge E^c
T^\gamma_{c \delta}  (\gamma_{ab})^{\a\b} \epsilon_{\rho\sigma} \iota_\alpha \iota_\beta
\left[ \iota_\gamma \delta(E^\rho)\wedge E^\delta \wedge \delta(E^\sigma) \right] \nonumber \\
&&=
 \epsilon^{abc} T^\gamma_{c \delta}  (\gamma_{ab})^{\a\b} \epsilon_{\rho\sigma}
\iota_\alpha \iota_\beta
\left[ \iota_\gamma \delta(E^\rho)\wedge E^\delta \wedge \delta(E^\sigma) \right] E^3 \nonumber \\
&&=
R \epsilon^{abc} (\gamma_c)^\gamma_{~\delta} (\gamma_{ab})^{\a\b}
\epsilon_{\rho\sigma} \iota_\alpha \iota_\beta
\left[ \iota_\gamma \delta(E^\rho)\wedge E^\delta \wedge \delta(E^\sigma) \right] E^3 \nonumber  = 0
\end{eqnarray}
where we have used $T^\gamma_{c \delta}  = {R} (\gamma^c)^\gamma_{~\delta}$, the fact that $\iota_\a \iota_\beta \iota_\gamma$
is totally symmetric with respect to the spinorial indices, and
the Fierz identity in $D=3$.

\subsection*{Relation between ${\mathbb Y}_{st}^{(0|2)}$ and
${\mathbb Y}_{susy}^{(0|2)}$}

There are two ways to compute the difference between  ${\mathbb Y}_{st}^{(0|2)}$ and
${\mathbb Y}_{susy}^{(0|2)}$. The first uses the fact that they are, from the mathematical point of view,
the Poincar\'e duals of embeddings of a submanifold ${\cal M}^{(3)}$ into
${\cal SM}^{(3|2)}$. Therefore, if the two embeddings gives two submanifolds in the same homology class
the corresponding Poincar\'e duals belongs to the same cohomology class. Thus, the difference is
$d$-exact.  The second way to verify this is to observe that the variation under a diffeomorphism $\xi$
(in the supermanifold) of the PCO is $d$-exact
\begin{eqnarray}
\label{reltA}
\delta {\mathbb Y}^{(0|2)} = {\cal L}_\xi {\mathbb Y}^{(0|2)} = d \Big( \iota_\xi  {\mathbb Y}^{(0|2)} \Big)
\end{eqnarray}
Therefore, we can relate two PCO's by infinitesimal changes of the background. With that
we can relate  ${\mathbb Y}_{susy}^{(0|2)}$ with the flat one
\begin{eqnarray}
\label{LCSH}
{\mathbb Y}^{(0|2)}_{susy/flat} = V^a\wedge V^b (\gamma_{ab})^{\a\b} \iota_\alpha \iota_\beta \delta^2(\psi)
\end{eqnarray}
where $V^a = dx^\a + \frac12 i \bar\theta \gamma^a d\theta$ and $\psi^\a = d\theta^\a$.

The flat Cartan-Maurer equations  immediately imply that $ dV^a = \frac i2 d\bar\theta \gamma^a d\theta$, $d\psi =0$ and therefore
\begin{eqnarray}
\label{LCSI}
d {\mathbb Y}^{(0|2)}_{ss/flat}  &=& i \bar\psi \gamma^a \psi \, \wedge V^b (\gamma_{ab})^{\a\b}  \iota_\alpha \iota_\beta \delta^2(\psi)
\nonumber \\ &=&
i  \iota_\alpha \iota_\beta (\bar\psi \gamma^a \psi) \wedge  V^b (\gamma_{ab})^{\a\b} \delta^2(\psi)
\nonumber \\
&=& 2 i \gamma^a_{\a\b} (\gamma_{ab})^{\a\b}  V^b \delta^2(\psi) = 0
\end{eqnarray}
It is manifestly invariant under supersymmetry, and  satisfies an interesting equation.
In $D=3$, with $N=1$ we have the following Chevalley-Eilenberg cohomology class
representative
\begin{eqnarray}
\label{LCSL}
\omega^{(3|0)} = \bar\psi \gamma_a \psi  V^a \,,
\end{eqnarray}
 which is supersymmetric (it is written in terms of supersymmetric variables) and is closed: $ d \omega^{(3|0)} =0$
 by using the Fierz identities. Now we can construct a $(3|2)$ form as
 follows
 \begin{eqnarray}
\label{LCSM}
{\mathbb Y}^{(0|2)}_{ss/flat}  \wedge \omega^{(3|0)} &=& V^a\wedge V^b (\gamma_{ab})^{\a\b} \iota_\alpha \iota_\beta \delta^2(\psi)\wedge  V^c \bar\psi \gamma_c \psi \nonumber \\
&=& \epsilon_{abc} V^a\wedge V^b \wedge V^c \epsilon_{\a\b} \delta(\psi^\alpha) \delta(\psi^\beta) = {\rm Vol}^{(3|2)}\,,
\end{eqnarray}
which is the volume form of the supermanifold ${\cal SM}^{(3|2)}$. In this sense, the PCO ${\mathbb Y}^{(0|2)}_{susy/flat}$ is the
Hodge dual to the Chevalley-Eilenberg cohomology class (\ref{LCSL}).
Expanding the flat bosonic vielbeins $V^a$ and using the derivative on the Dirac delta functions,
we can rewrite ${\mathbb Y}^{(0|2)}_{susy/flat}$ as
follows
\begin{eqnarray}
\label{LCSN}
{\mathbb Y}^{(0|2)}_{susy/flat} = d \Big[ \Omega^{(-1|2)} \Big] + {\mathbb Y}^{(0|2)}_{st}
\end{eqnarray}
where $ \Omega^{(-1|2)}$ is a $(-1|2)$ form in the space of integral forms. This proves that
the difference between the supersymmetric flat PCO and the spacetime PCO is an exact term. The difference between
flat and curved supersymmetric PCO's is again $d$-exact (since it is produced via a diffeomorphism), so that
${\mathbb Y}_{st}^{(0|2)}$ and
${\mathbb Y}_{susy}^{(0|2)}$
indeed belong to the same cohomology class.

\end{document}